\documentclass[sigconf,authorversion,nonacm]{acmart}
\AtBeginDocument{%
  \providecommand\BibTeX{{%
    \normalfont B\kern-0.5em{\scshape i\kern-0.25em b}\kern-0.8em\TeX}}}

\copyrightyear{2024} 
\acmYear{2024} 
\setcopyright{acmlicensed}\acmConference[ETRA '24]{2024 Symposium on Eye Tracking Research and Applications}{June 4--7, 2024}{Glasgow, United Kingdom}
\acmBooktitle{2024 Symposium on Eye Tracking Research and Applications (ETRA '24), June 4--7, 2024, Glasgow, United Kingdom}
\acmDOI{10.1145/3649902.3656495}
\acmISBN{979-8-4007-0607-3/24/06}

\acmConference[ETRA '24]{Make sure to enter the correct
  conference title from your rights confirmation emai}{June 04--07,
  2024}{Glasgow, United Kingdom}
\acmBooktitle{ACM Symposium on Eye Tracking Research \& Applications (ETRA),
 June 04--07, 2024, Glasgow, United Kingdom} 
\acmPrice{15.00}
\acmISBN{978-1-4503-XXXX-X/18/06}

\usepackage{cleveref}
\usepackage{multirow}
\usepackage{makecell}
\usepackage{subcaption}
\usepackage{enumitem}
\usepackage{stfloats}

\begin{document}

\title{EyeLiveMetrics: Real-time Analysis of Online Reading with Eye Tracking}


\author{Daniel Hienert}
\email{daniel.hienert@gesis.org}
\orcid{0000-0002-2388-4609}
\affiliation{%
 \institution{GESIS - Leibniz Institute for the Social Sciences}
 \streetaddress{Unter Sachsenhausen 6-8}
 \city{Cologne}
 \country{Germany}
}
\author{Heiko Schmidt}
\email{heiko.schmidt@gesis.org}
\orcid{0009-0009-7795-7592}
\affiliation{%
 \institution{GESIS - Leibniz Institute for the Social Sciences}
 \streetaddress{Unter Sachsenhausen 6-8}
 \city{Cologne}
 \country{Germany}
}
\author{Thomas Krämer}
\email{thomas.kraemer@gesis.org}
\orcid{0000-0003-0507-7843}
\affiliation{%
 \institution{GESIS - Leibniz Institute for the Social Sciences}
 \streetaddress{Unter Sachsenhausen 6-8}
 \city{Cologne}
 \country{Germany}
}
\author{Dagmar Kern}
\email{dagmar.kern@gesis.org}
\orcid{0000-0003-1794-625X}
\affiliation{%
 \institution{GESIS - Leibniz Institute for the Social Sciences}
 \streetaddress{Unter Sachsenhausen 6-8}
 \city{Cologne}
 \country{Germany}
}

\renewcommand{\shortauthors}{Hienert et al.}

\begin{abstract}

Existing eye tracking software have certain limitations, especially with respect to monitoring reading online: (1) Most eye tracking software record eye tracking data as raw coordinates and stimuli as screen images/videos, but without inherent links between both. Analysts must draw areas of interest (AOIs) on webpage text for more fine-grained reading analysis. (2) The computation and analysis of fixation and reading metrics are done after the experiment and thus cannot be used for live applications.

We present EyeLiveMetrics, a browser plugin that automatically maps raw gaze coordinates to text in real time. The plugin instantly calculates, stores, and provides fixation, saccade, and reading measures on words and paragraphs so that gaze behavior can be analyzed immediately. We also discuss the results of a comparative evaluation. EyeLiveMetrics offers a flexible way to measure reading on the web - for research experiments and live applications.

\end{abstract}




\maketitle

\section{Introduction}
Eye tracking is a well-established method for measuring and analyzing the eye movements of individuals in various domains, such as marketing \cite{wedel2017review}, human-computer interaction (HCI) and usability \cite{poole2006eye} as well as reading research \cite{just1980theory}. This is partly because eye movements are good indicators of the user's attention, and a lot can be inferred about the user's cognitive state due to the correlation between eye movements and cognitive processes \cite{rayner1998eye}. This makes eye tracking beneficial in experimental user studies to research human behavior. However, most eye tracking systems lack the generic functionality that maps raw eye tracking data to the visible stimuli. The mapping is usually done in the time-consuming process of drawing areas of interest (AOIs) by hand on each relevant screenshot. With the increasing interest in using eye tracking for reading-related activities on the web such as information retrieval \cite{Buscher2012}, online news consumption \cite{Bhattacharya2020}, fake news detection \cite{Bozkir2022} or social media usage \cite{vraga2016beyond} this workflow is not efficient. Drawing AOIs for hundreds of web pages, each with multiple AOIs visited by several participants, is a major time commitment.

With the EyeMetricsLive plugin, we want to provide a solution for web stimuli in the form of a browser plugin that (1) captures eye tracking data live, (2) maps them to text on a web page even down to the word level, (3) calculates eye tracking metrics instantly during recording. The plugin enables researchers to analyze gaze data collected in experimental user studies efficiently and developers to adapt their interfaces based on previous gaze behavior, similar to the approach introduced as "The text 2.0 framework" by Bieder et al. \cite{Biedert2010}. For example, a web page can respond to reading comprehension difficulties by automatically switching to easy language, or a search engine can suggest search terms or re-rank search results based on what the user has read so far. Furthermore, the plugin eliminates the need for the time-consuming manual drawing of AOIs in eye tracking experiments, as it automatically maps gaze data to HTML elements. EyeLiveMetrics is an open-source modular plugin compatible with any eye tracker as long as it has access to the eye tracking data stream.

In this paper, we introduce EyeLiveMetrics, its processes and the available metrics. Furthermore, we give insights into the validity of EyeLiveMetrics. 

\section{Related Work}
Eye tracking has a long tradition in reading research and is considered a reliable indicator for individuals’ moment-by-moment cognitive processing \cite{just1980theory, rayner1998eye}. Various reading measures have been devised to capture these processes mainly based on fixation and saccades measures, like first-pass fixation duration, first-pass regression or re-reading duration. A common application of these reading measures is to classify whether a reader is reading a text in depth or skimming \cite{biedert2012robust, Kelton2019, Bhattacharya2020, Gwizdka2014}. Although several studies have demonstrated the feasibility of this classification task, so far, no system or prototype has been developed to perform it in real time. Another promising application is the detection of reading comprehension, e.g., \cite{just2018using, meziere2023using, Ahn2020} in general and, more specifically, in the field of fake news detection on the web. Research by Bozkir et al. \cite{Bozkir2022} and Sümer et al. \cite{sumer2021fakenewsperception} has shown that different eye movements are associated with reading news articles with true and false information. 

In the context of live applications, Biedert et al. introduced the Text 2.0 framework \cite{Biedert2010}, which allows us to react to reading behavior in real time. In the Eye Book \cite{biedert2009eye}, the user's gaze behavior is used to determine which text segments are currently being read, and based on this, additional effects such as sounds or images are generated and presented to the user. Another real-time approach in Interactive Information Retrieval uses gaze data to suggest query reformulations in a search engine prototype \cite{Eickhoff2015}. In both examples, the system knows what the user is currently reading or has read before.

Over the years, several software tools have been developed to support researchers in conducting eye tracking experiments and analyze gaze data. Some commercial tools, such as Tobii Pro Lab, support reading measures\footnote{https://connect.tobii.com/s/article/How-reading-metrics-work?language=en\_US} for specific text created by the experimenter with an integrated editor. The tools then automatically generate AOIs for characters, words, and sentences. However, for other types of stimuli, such as web studies, analysts still need to draw AOIs manually. EyeMap \cite{tang2012eyemap} and the ReadingProtocol \cite{Hienert2019} are open-source tools that overcome this drawback by mapping the gaze data to automatically generated AOIs based on HTML elements of the web page. Based on them, research prototypes have been built, such as WebgazeAnalyzer \cite{Beymer2005WEbGazeAnalyzer}, WebEyeMapper \cite{Reeder2001} and ReMA \cite{Valdunciel2022}, but none of them considers real-time usage. Webgazer.js \cite{papoutsaki2016webgazer} allows the integration of eye movement data into a live system. However, it doesn't take into account the underlying content. It estimates eye tracking data based on webcam input and current interaction behavior on the screen, which comes with a loss of accuracy. Therefore, it is not yet applicable to use cases where high accuracy is necessary. This is where the EyeLiveMetrics plugin comes in. It combines highly accurate, real-time eye tracking with automatic AOI detection.

\section{EyeLiveMetrics - The Browser plugin}
In this section, we present the implementation of the open-source browser plugin EyeLiveMetrics\footnote{\label{footnote:link_GIT}EyeLiveMetrics is an open-source software under GPLv3 licence at https://git.gesis.org/iir/eyelivemetrics}. The main goals of this plugin are (1) eliminating the time-consuming process of mapping eye tracking data to text with AOIs, (2) computing fixation-, saccades- or reading measures for text on web pages on the fly, (3) enabling real-time analysis of reading and therewith immediate reaction to detected gaze behavior in an interactive setting. In the following, we give an overview of how the plugin collects the eye tracking data and the data processing procedure, which includes the mapping from gaze data to the text of the web page and the computation of the eye tracking metrics.

\subsection{Gaze Data Collection}
\label{chp:gaze_data_flow}
EyeLiveMetrics is built upon the Reading Protocol tool \cite{Hienert2019}, which uses raw eye tracking data and stored HTML pages to compute fixations on words in a rendering process after the user experiment. Thereby, coordinates were mapped to the bounding boxes of individual words, representing the AOIs, and later stored as word-eye-fixations in a database. This process can be error-prone for experiments on the web with arbitrary web pages. The Reading Protocol only stores HTML pages once loaded, but many web pages are dynamic, reloading content via AJAX calls or having dynamic content items such as menus, drop-down boxes, or carousels. Using a web page that is only stored once after the initial loading for the mapping process would not reflect the state over the entire time frame that a subject reads it. Accordingly, this can lead to errors in the mapping process from eye tracking coordinates to AOIs.

\begin{figure*}
    \centering
    \includegraphics[width=0.8\textwidth]{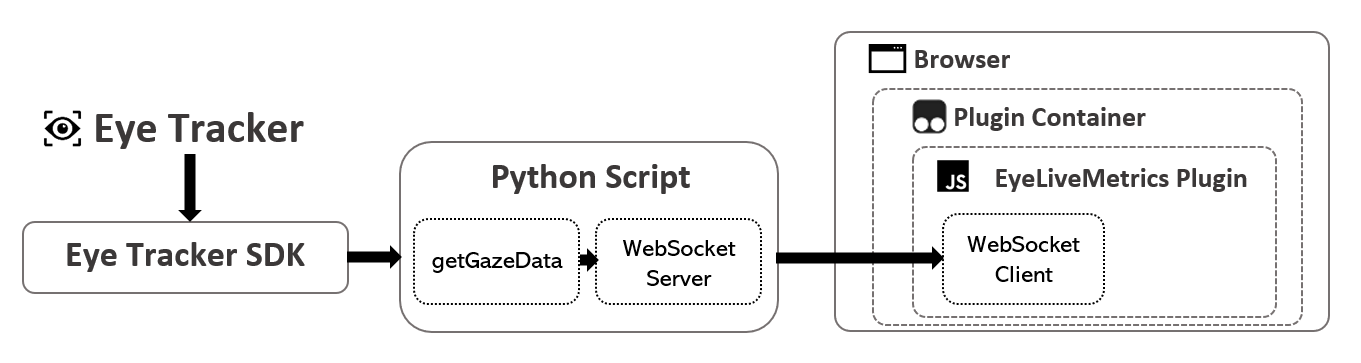}
    \caption{Flow of gaze data from eye tracker to the EyeLiveMetrics plugin}
    \label{fig:gaze_data_flow}
\end{figure*}

EyeLiveMetrics performs the mapping process in real time. Figure \ref{fig:gaze_data_flow} gives an overview of the flow of gaze data. We implemented a Python script to transfer eye tracking coordinates to the plugin. The script accesses the eye tracker with, in our case, the Tobii Pro SDK for Python\footnote{https://developer.tobiipro.com/python/python-getting-started.html}. The SDK provides a continuous stream of the two-dimensional display coordinates (x,y) with timestamps and the 3D space coordinates (x,y,z), gaze-origin and gaze-position in the user coordinate system. The Python script also acts as a WebSocket server and sends the coordinates to the plugin, which acts as a client. 

The plugin itself is implemented as a Tampermonkey\footnote{https://www.tampermonkey.net/} user script. Tampermonkey is a plugin container that allows the execution of arbitrary JavaScript code on any web page. It supports many browsers, such as Chrome, Edge, Safari, Opera, or Firefox. It allows a pattern-based inclusion or exclusion of web pages for which the user script is executed.

\subsection{Data Processing Procedure}
\label{chp:plugin_processing_steps}

\begin{figure*}
    \centering
    \includegraphics[width=0.8\textwidth]{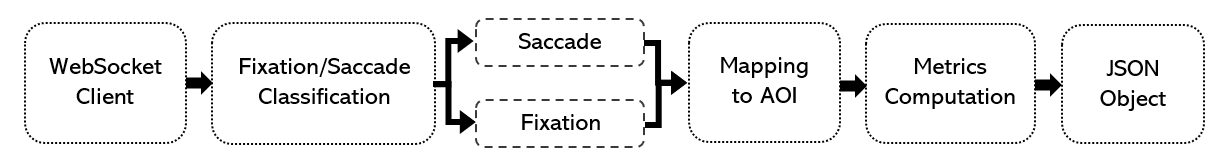}
    \caption{Data processing in the EyeLiveMetrics plugin}
    \label{fig:data_processing}
\end{figure*}

Figure \ref{fig:data_processing} shows the data processing procedure in the EyeLiveMetrics plugin. Eye tracking coordinates are received in the WebSocket client. It uses a Velocity-Threshold Identification (I-VT) algorithm \cite{SalvucciGoldberg2000} to classify raw gaze points as fixations or saccades based on the eye's directional shift velocity. Input parameters are gaze-origin and gaze-position coordinates to compute the gaze vector and compare it with previous ones. It returns the classification of fixation or saccade based on an adjustable threshold, e.g., 30° per second. For gaze points classified as a saccade, we store their position and velocity. If classification changes to fixation, the final saccade is computed and mapped to an AOI. AOIs used for saccade mapping are text paragraphs, images, and videos. If the start- and endpoint lies within an AOI, the saccade is saved for this AOI. At last, we compute the saccade duration and length. In Section \ref{chp:et_metrics}, we will describe the saccade metrics in more detail.

When the classifier identifies a fixation, a fixation object is created. It includes properties such as start- and end time, the mapped HTML element (word, image, video) as AOI, the object bounding box, and object-specific metrics (e.g., word index/offset in the text node). When the fixation ends, fixation and reading metrics are computed. We refer to the next Section \ref{chp:et_metrics} for a detailed overview. All data is stored locally in the browser and then every five seconds or when the status of the browser tab changes in a database, e.g., on tab close or change. In general, we store participant ID, stimulus (URL of the web page), start-, end time, fixations, and saccades objects for words, images, and videos as well as the whole web page as text, a word and a sentence index for the whole web page text.

\subsection{Metrics for Reading}
\label{chp:et_metrics}

The metrics of EyeLiveMetrics follow the implementation of fixation\textendash, saccade-, and reading metrics by Tobii Pro Lab\footnote{https://connect.tobii.com/s/article/understanding-tobii-pro-lab-eye-tracking-metrics?language=en\_US}. A full list of all available metrics can be found in the GIT repository\footref{footnote:link_GIT} of EyeLiveMetrics.

Fixations are the prolonged retention of the gaze on one point, for textual stimuli on words to understand their syntactic and semantic meaning \cite{reichle2012using}. EyeLiveMetrics stores fixation measures for each web page element, such as words. First, we have the common eye tracking metrics such as total fixation duration, number of fixations, timestamp of the first fixation, time to first fixation, first fixation duration, and average fixation duration. Further, we have measures specific to words, e.g., character index and sentence number in HTML documents. At last, we also have timestamps, fixation group numbers, and stimuli bounding boxes.

Saccades are rapid movements of the gaze between fixations. For text stimuli, they describe the movement from word to word. This means that saccades can abstractly describe the reading behavior on text. For example, to understand the scan path on paragraphs \cite{Valdunciel2022} or for the distinction between reading and skimming \cite{biedert2012robust}. EyeLiveMetrics stores saccades for textual paragraphs. For each saccade, it stores the timestamps of the entry- and exit saccade, its sequential index, its duration, its length, its amplitude, its peak velocity, and its 2-dimensional direction vector. 

Reading measures can describe additional characteristics of reading behavior, such as analyzing reading comprehension \cite{meziere2023using}. In case the HTML element is a word, EyeLifeMetrics stores the following reading measures: First-pass first fixation duration, first-pass fixation group number, first-pass regression, first-pass duration, regression path duration, selective regression path duration, and re-reading duration.

\section{Usage of EyeLiveMetrics}

Once the EyeLiveMetrics plugin has been added to the browser and the other necessary resources (e.g. database and web server) have been set up according to the instructions on the project website, the setup is ready to be used either in a user study or in an interactive environment. In both cases, the eye tracker must be calibrated to the user's eyes using the calibration tool usually available in the eye tracker management software. Next, the browser can be opened, and the EyeLiveMetrics plugin can be activated. From now on, gaze data collection (see \ref{chp:gaze_data_flow}) will take place for each web page visited.

\begin{figure*}
    \centering
    \includegraphics[width=0.8\textwidth]
    {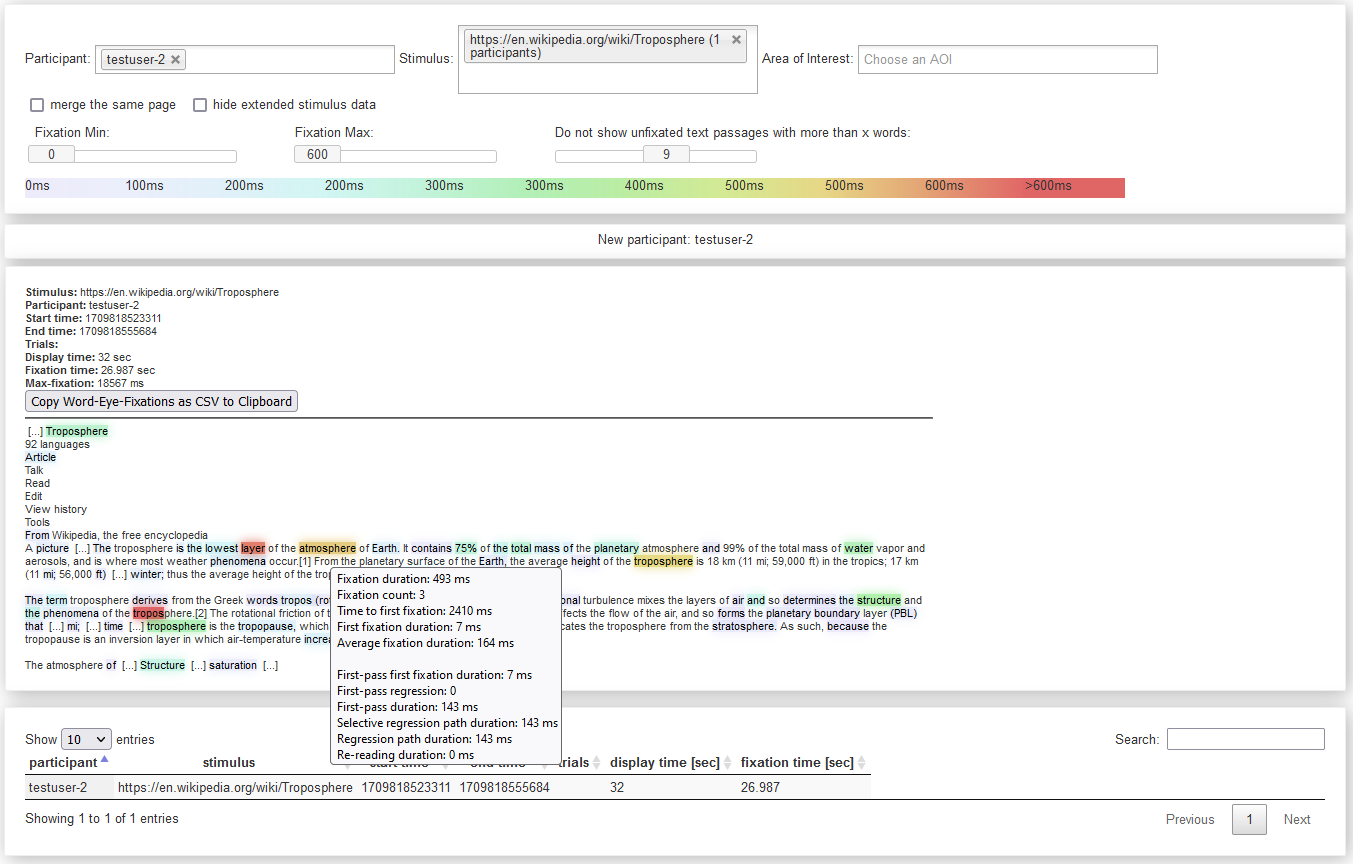}
    \caption{EyeLiveMetric's user interface showing analysis results adapted from the reading protocol UI \cite{Hienert2019}.}
    \label{fig:analysisUI}
\end{figure*}

After the data collection in user studies, the experimenter can immediately see the analysis results by simply launching the analysis user interface (see Figure \ref{fig:analysisUI}). This also allows the data to be used directly in retrospective interviews. Fixation-, saccades-, and reading metrics can be seen in a pop-up window for each word. 

\begin{figure*}
    \centering
    \includegraphics[width=0.8\textwidth]
    {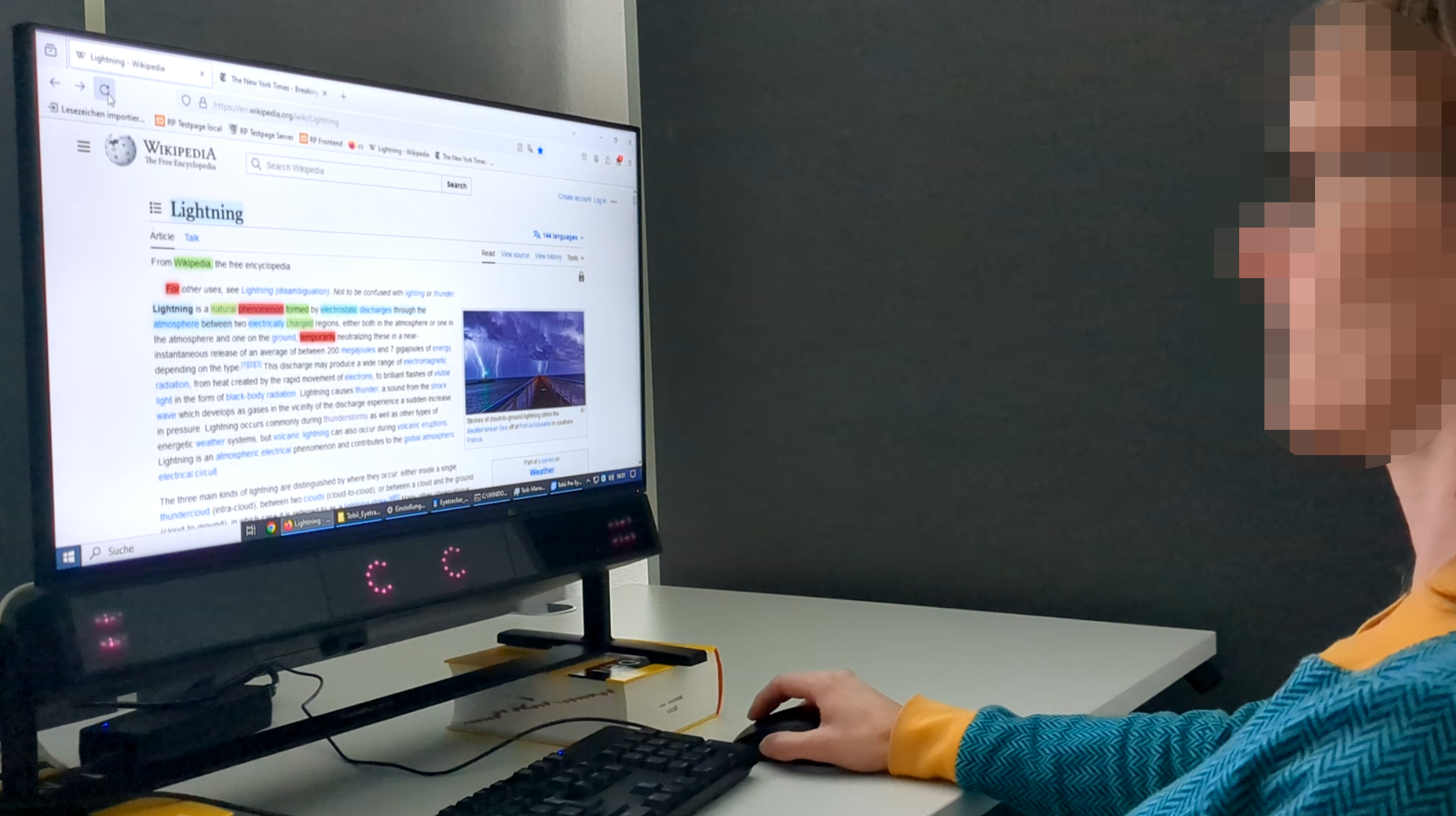}
    \caption{Participant reading a Wikipedia article and generating a heat map on the fly.}
    \label{fig:wikipedia}
\end{figure*}

In real-time applications, the analyzed data can be used immediately to manipulate the interface. For example, Figure \ref{fig:wikipedia} shows how the intensity of reading a word can be shown directly to the reader. Here, the word's fixation duration determines its background color. Although the usefulness of this approach is questionable, it illustrates that it is possible to generate a heat map on the fly.  

\section{Evaluation}
We evaluated EyeLiveMetrics in two different ways. First, we examined the performance of EyeLiveMetrics and then compared the results with those of Tobii Pro Lab for the same text stimulus. We begin this section with a description of the technical environment we used in the two experiments and present both evaluations afterward. 

\subsection{Technical evaluation environment}
\label{chp:exp_technical_specs}

For the evaluation, we used a Tobii Pro Spectrum device with a sampling rate of 300 Hz. The sampling rate is an upper limit of how much time a live application such as EyeLiveMetrics can use to fully process a single gaze coordinate. Both experiments were conducted on a PC with Intel i7-6700 CPU@3.40GHz and 16GB RAM with Firefox V117.0 as a web browser. As mentioned in \Cref{chp:gaze_data_flow}, we integrate EyeLiveMetrics as a Tampermonkey plugin.

\subsection{Experiment: EyeLiveMetrics Performance}
\label{chp:eval_perf}
Since the EyeLiveMetrics is a browser plugin that computes eye tracking metrics in real time (cf. \Cref{chp:et_metrics}), its main loop has to run at least as fast as the eye tracker records and outputs raw gaze coordinates according to its sampling rate. For each coordinate, the plugin has to receive the coordinates, classify them into saccades or fixations, then map them to an HTML element (text, word, image, video), and at last, compute the fixation-, saccade- and reading metrics. Therefore, each iteration of the plugin has to be quick enough to handle the next raw gaze data point.

We conducted a performance test with the technical specifications described in \Cref{chp:exp_technical_specs} and recorded ten minutes of reading a large article on Wikipedia. The plugin recorded and processed n=176,091 raw gaze coordinates with an average computation time of 0.464ms (SD=1.037) for the iteration of a single raw gaze coordinate. This processing time is fast enough for eye trackers with frequencies up to 1200Hz (new raw gaze coordinate each 0.833ms).

\subsection{Experiment: Metrics Accuracy}
\label{experiment_metric_accuracy}

\renewcommand\arraystretch{1.3}

\begin{center}
    \begin{table*}[tp]
    \scriptsize
        \centering
        \caption{Comparison of eye tracking metrics of Tobii Pro Lab and EyeLiveMetrics. All time measures are in ms.}
        \label{tab:comp_ETmetrics}
        \resizebox{\textwidth}{!}{
        \begin{tabular}{|c|c|c|c|c|c|c|c|}
        \hline
             \multirow{2}{*}{Metrics} & \multicolumn{7}{c|}{Fixation Metrics} \\ 
             & TFD & AFD & MiFD & MaFD & \#F & TFF & FFD \\ \Xhline{3\arrayrulewidth}
             Pearson & 
             $1.0$ & $0.99$ & $0.996$ & $0.962$ &
             $0.999$ & $1.0$ & $0.973$ \\ \hline
             MAE + std & 
             $5.364\pm 6.067$ & $0.734\pm 0.675$ & $0.281\pm 0.119$ & $3.879\pm 7.542$ &
             $1.0\pm 1.128$ & $3.595\pm 2.297$ & $0.636\pm 1.327$ \\ \Xhline{5\arrayrulewidth}
             & \multicolumn{6}{c|}{Reading Metrics} & \multicolumn{1}{c}{} \\
             & FpFFD & FpD & FpR & RPD & sRPD & RRD & \multicolumn{1}{c}{} \\ \Xcline{1-7}{3\arrayrulewidth}
             Pearson & 
             $0.979$ & $1.0$ & $1.0$ & $1.0$ &
             $1.0$ & $1.0$ &  \multicolumn{1}{c}{} \\ \cline{1-7}
             MAE + std &
             $0.495\pm 1.226$ & $3.757\pm 6.006$ & $0.0\pm 0.0$ &
             $4.393\pm 6.446$ & $4.353\pm 5.884$ & $1.021\pm 3.24$ & \multicolumn{1}{c}{} \\ \cline{1-7}
        \end{tabular}
        }
    \end{table*}
\end{center}

\renewcommand\arraystretch{1}

EyeLiveMetrics can only be a valuable eye tracking software tool when its computed eye tracking metrics are accurate and precise. To test that condition, we compare EyeLiveMetrics results to those from Tobii Pro Lab. Therefore, we created a reading experiment ('text stimulus') in Tobii Pro Lab, which is the only stimulus that provides reading metrics. With the internal text editor, we created a text paragraph with the font Arial, a line height of 1.2 and a font size of 48px. AOIs for words are automatically created by the editor. We then recorded a participant reading the text paragraph. After the experiment, raw eye tracking data was exported via the 'data export' functionality. Additionally, we exported fixation- and reading metrics for all words in the stimulus from Tobii Pro Lab as a CSV file via the 'metrics export' functionality. We took a screenshot of the Tobii text stimulus and recreated it as a web page using HTML and CSS. Using a Python script, we load the raw eye-tracking data, replay it and send it to the browser, which displays the reconstructed text stimulus. The EyeLiveMetrics plugin then processes the eye tracking data, maps them to words, and computes metrics. All metrics are stored in a database. After exporting the eye tracking metrics from both software, Tobii Pro Lab and EyeLiveMetrics, we compare them for the following eye tracking metrics:

 \begin{itemize}[leftmargin=10pt, rightmargin=10pt, noitemsep,topsep=10pt]
    \item[] \textbf{Fixations:} Total Fixation Duration (TFD), Average Fixation Duration (AFD), Minimum Fixation Duration (MiFD), Maximum Fixation Duration (MaFD), Number of Fixations (\#F), Timestamp of First Fixation (TFF) and First Fixation Duration (FFD)
    \item[] \textbf{Reading:} First-Pass First Fixation Duration (FpFFD), First-Pass Duration (FpD), First-pass Regression (FpR), Regression-Path Duration (RPD), Selective Regression-Path Duration (sRPD) and Re-Reading Duration (RRD)
\end{itemize}
 
\Cref{tab:comp_ETmetrics} shows the comparison results. We report Pearson correlation ($\rho$) and Mean Absolute Error (MAE) for each fixation- and reading metric. Overall, the fixation metrics have a very high confidence level of similarity between EyeLiveMetrics and Tobii ($0.962\leq\rho\leq1.0$) and the MAE on average, has an error margin of $<6$ ms. For the reading metrics, we can report similar results ($0.979\leq\rho\leq 1.0$). 

We faced several challenges while comparing fixations on both text stimuli (Tobii Pro Lab \& web page): (1) Although using the same font, the same font size, and the same line height, we observed a slightly different rendering of the text in Tobii's internal view and for the web browser. We needed to adapt the line height via CSS to fit Tobii's word bounding boxes. (2) Fixations on spaces between words in Tobii are split into the previous and the next word in Tobii. Experimentally, we found a split of 1/3 of the space to the previous word and 2/3 to the next and implemented that in our code. (3) To have a solid basis for comparison, we have set the fixation filters as simple as possible. Both sources use an I-VT filter with 30 degrees/sec as a threshold. For Tobii, we set the velocity calculator to 3ms so that only the last two coordinates are used for the I-VT calculation. We have switched off all other filters. (4) There are differences in the algorithm between Tobii and EyeLiveMetrics due to the live vs. post-experiment character. For example, EyeLiveMetrics can only evaluate current live coordinates within a queue. This can affect the filter implementation, e.g. for noise filters or the velocity calculator.

Nonetheless, we found a very strong correlation between Tobii's and our fixation and reading metrics, proving that the metrics implementation works accurately and in real time.

\section{Conclusion}

We presented EyeLiveMetrics, an open-source browser plugin to capture gaze data in real time in web environments. Eye coordinates, derived fixations, and saccades are directly mapped to text on webpages. Reading measures are computed live for fixated words, sentences, and texts. This is in contrast to existing eye tracking analysis software that often only allows post-analysis with manual mapping of AOIs and for a limited set of measures. We have shown that the plugin works efficiently for eye trackers with high sampling rates and computes all measures with a high correlation to Tobii Pro Lab measures, as a representative of commercial eye tracking software. 
Real-time analysis of online reading supports experiments in which participants search, browse, and learn information online. Experimenter have instant access to fixation-, saccade- and reading metrics without the burden of AOI mapping and metrics computation. This additionally allows for live applications that react, for example, to longer fixations or reading difficulties.

\begin{acks}
This work was partly funded by the DFG, grant no. 525041402; the ECHOES project at GESIS.
\end{acks}

\bibliographystyle{ACM-Reference-Format}
\bibliography{ETRA24EyeLiveMetrics}

@inproceedings{SalvucciGoldberg2000,
author = {Salvucci, Dario D. and Goldberg, Joseph H.},
title = {Identifying Fixations and Saccades in Eye-Tracking Protocols},
year = {2000},
isbn = {1581132808},
publisher = {Association for Computing Machinery},
address = {New York, NY, USA},
url = {https://doi.org/10.1145/355017.355028},
doi = {10.1145/355017.355028},
booktitle = {Proceedings of the 2000 Symposium on Eye Tracking Research \& Applications},
pages = {71–78},
numpages = {8},
location = {Palm Beach Gardens, Florida, USA},
series = {ETRA '00}
}

@inproceedings{Hienert2019,
author = {Hienert, Daniel and Kern, Dagmar and Mitsui, Matthew and Shah, Chirag and Belkin, Nicholas J.},
title = {Reading Protocol: Understanding What Has Been Read in Interactive Information Retrieval Tasks},
year = {2019},
isbn = {9781450360258},
publisher = {Association for Computing Machinery},
address = {New York, NY, USA},
url = {https://doi.org/10.1145/3295750.3298921},
doi = {10.1145/3295750.3298921},
booktitle = {Proceedings of the 2019 Conference on Human Information Interaction and Retrieval},
pages = {73–81},
numpages = {9},
location = {Glasgow, Scotland UK},
series = {CHIIR '19}
}

@article{tang2012eyemap,
  title={EyeMap: a software system for visualizing and analyzing eye movement data in reading},
  author={Tang, Siliang and Reilly, Ronan G and Vorstius, Christian},
  journal={Behavior research methods},
  volume={44},
  pages={420--438},
  year={2012},
  publisher={Springer},
  doi={https://doi.org/10.3758/s13428-011-0156-y}
}

@inproceedings{Beymer2005WEbGazeAnalyzer,
author = {Beymer, David and Russell, Daniel M.},
title = {WebGazeAnalyzer: A System for Capturing and Analyzing Web Reading Behavior Using Eye Gaze},
year = {2005},
isbn = {1595930027},
publisher = {Association for Computing Machinery},
address = {New York, NY, USA},
url = {https://doi.org/10.1145/1056808.1057055},
doi = {10.1145/1056808.1057055},
booktitle = {CHI '05 Extended Abstracts on Human Factors in Computing Systems},
pages = {1913–1916},
numpages = {4},
location = {Portland, OR, USA},
series = {CHI EA '05}
}

@inproceedings{Valdunciel2022,
author = {Valdunciel, Pablo and Bhatti, Omair Shahzad and Barz, Michael and Sonntag, Daniel},
title = {Interactive Assessment Tool for Gaze-Based Machine Learning Models in Information Retrieval},
year = {2022},
isbn = {9781450391863},
publisher = {Association for Computing Machinery},
address = {New York, NY, USA},
url = {https://doi.org/10.1145/3498366.3505834},
doi = {10.1145/3498366.3505834},
booktitle = {Proceedings of the 2022 Conference on Human Information Interaction and Retrieval},
pages = {332–336},
numpages = {5},
location = {Regensburg, Germany},
series = {CHIIR '22}
}

@inproceedings{Reeder2001,
author = {Reeder, Robert W. and Pirolli, Peter and Card, Stuart K.},
title = {WebEyeMapper and WebLogger: Tools for Analyzing Eye Tracking Data Collected in Web-Use Studies},
year = {2001},
isbn = {1581133405},
publisher = {Association for Computing Machinery},
address = {New York, NY, USA},
url = {https://doi.org/10.1145/634067.634082},
doi = {10.1145/634067.634082},
booktitle = {CHI '01 Extended Abstracts on Human Factors in Computing Systems},
pages = {19–20},
numpages = {2},
location = {Seattle, Washington},
series = {CHI EA '01}
}

@inproceedings{Kelton2019,
author = {Kelton, Conor and Wei, Zijun and Ahn, Seoyoung and Balasubramanian, Aruna and Das, Samir R. and Samaras, Dimitris and Zelinsky, Gregory},
title = {Reading Detection in Real-Time},
year = {2019},
isbn = {9781450367097},
publisher = {Association for Computing Machinery},
address = {New York, NY, USA},
url = {https://doi.org/10.1145/3314111.3319916},
doi = {10.1145/3314111.3319916},
booktitle = {Proceedings of the 11th ACM Symposium on Eye Tracking Research \& Applications},
articleno = {43},
numpages = {5},
location = {Denver, Colorado},
series = {ETRA '19}
}

@inproceedings{Ahn2020,
author = {Ahn, Seoyoung and Kelton, Conor and Balasubramanian, Aruna and Zelinsky, Greg},
title = {Towards Predicting Reading Comprehension From Gaze Behavior},
year = {2020},
isbn = {9781450371346},
publisher = {Association for Computing Machinery},
address = {New York, NY, USA},
url = {https://doi.org/10.1145/3379156.3391335},
doi = {10.1145/3379156.3391335},
booktitle = {ACM Symposium on Eye Tracking Research and Applications},
articleno = {32},
numpages = {5},
location = {Stuttgart, Germany},
series = {ETRA '20 Short Papers}
}

@inproceedings{Bhattacharya2020,
author = {Bhattacharya, Nilavra and Rakshit, Somnath and Gwizdka, Jacek},
title = {Towards Real-Time Webpage Relevance Prediction UsingConvex Hull Based Eye-Tracking Features},
year = {2020},
isbn = {9781450371353},
publisher = {Association for Computing Machinery},
address = {New York, NY, USA},
url = {https://doi.org/10.1145/3379157.3391302},
doi = {10.1145/3379157.3391302},
booktitle = {ACM Symposium on Eye Tracking Research and Applications},
articleno = {28},
numpages = {10},
location = {Stuttgart, Germany},
series = {ETRA '20 Adjunct}
}

@inproceedings{Bozkir2022,
author = {Bozkir, Efe and Kasneci, Gjergji and Utz, Sonja and Kasneci, Enkelejda},
title = {Regressive Saccadic Eye Movements on Fake News},
year = {2022},
isbn = {9781450392525},
publisher = {Association for Computing Machinery},
address = {New York, NY, USA},
url = {https://doi.org/10.1145/3517031.3529619},
doi = {10.1145/3517031.3529619},
booktitle = {2022 Symposium on Eye Tracking Research and Applications},
articleno = {7},
numpages = {7},
location = {Seattle, WA, USA},
series = {ETRA '22}
}

@article{biedert2009eye,
author = "Biedert, Ralf and Buscher, Georg and Dengel, Andreas",
title = "The eyeBook – Using Eye Tracking to Enhance the Reading Experience",
year = 2010,
journal = "Informatik-Spektrum",
volume = "33",
number = "3",
publisher = "Springer-Verlag",
address = "Berlin Heidelberg",
issn = "1432-122X",
pages = "272--281",
}

@article{Buscher2012,
author = {Buscher, Georg and Dengel, Andreas and Biedert, Ralf and Elst, Ludger V.},
title = {Attentive Documents: Eye Tracking as Implicit Feedback for Information Retrieval and Beyond},
year = {2012},
issue_date = {January 2012},
publisher = {Association for Computing Machinery},
address = {New York, NY, USA},
volume = {1},
number = {2},
issn = {2160-6455},
url = {https://doi.org/10.1145/2070719.2070722},
doi = {10.1145/2070719.2070722},
journal = {ACM Trans. Interact. Intell. Syst.},
month = {jan},
articleno = {9},
numpages = {30}
}

@inproceedings{Gwizdka2014,
author = {Gwizdka, Jacek},
title = {Characterizing Relevance with Eye-Tracking Measures},
year = {2014},
isbn = {9781450329767},
publisher = {Association for Computing Machinery},
address = {New York, NY, USA},
url = {https://doi.org/10.1145/2637002.2637011},
doi = {10.1145/2637002.2637011},
booktitle = {Proceedings of the 5th Information Interaction in Context Symposium},
pages = {58–67},
numpages = {10},
location = {Regensburg, Germany},
series = {IIiX '14}
}

@inproceedings{Eickhoff2015,
author = {Eickhoff, Carsten and Dungs, Sebastian and Tran, Vu},
title = {An Eye-Tracking Study of Query Reformulation},
year = {2015},
isbn = {9781450336215},
publisher = {Association for Computing Machinery},
address = {New York, NY, USA},
url = {https://doi.org/10.1145/2766462.2767703},
doi = {10.1145/2766462.2767703},
booktitle = {Proceedings of the 38th International ACM SIGIR Conference on Research and Development in Information Retrieval},
pages = {13–22},
numpages = {10},
location = {Santiago, Chile},
series = {SIGIR '15}
}

@article{wedel2017review,
  title={A review of eye-tracking research in marketing},
  author={Wedel, Michel and Pieters, Rik},
  journal={Review of Marketing Research},
  volume={4},
  pages={123--147},
  year={2017},
  publisher={Emerald Publishing Group},
  doi={https://doi.org/10.1108/S1548-6435(2008)0000004009}
}

@incollection{poole2006eye,
  title={Eye Tracking in HCI and Usability Research},
  author={Poole, Alex and Ball, Linden J},
  booktitle={Encyclopedia of human computer interaction},
  pages={211--219},
  year={2006},
  editor={Claude Ghaoui},
  publisher={IGI global},
  doi={10.4018/978-1-59140-562-7.ch034}
}

@article{just1980theory,
  title={A theory of reading: from eye fixations to comprehension},
  author={Just, Marcel A and Carpenter, Patricia A},
  journal={Psychological review},
  volume={87},
  number={4},
  pages={329},
  year={1980},
  publisher={American Psychological Association},
  url={https://psycnet.apa.org/doi/10.1037/0033-295X.87.4.329}
}

@article{vraga2016beyond,
  author = {Emily Vraga, Leticia Bode and Sonya Troller-Renfree},
title = {Beyond Self-Reports: Using Eye Tracking to Measure Topic and Style Differences in Attention to Social Media Content},
journal = {Communication Methods and Measures},
volume = {10},
number = {2-3},
pages = {149-164},
year = {2016},
publisher = {Routledge},
doi = {10.1080/19312458.2016.1150443},
URL = {https://doi.org/10.1080/19312458.2016.1150443},
}

@inproceedings{biedert2012robust,
  author = {Biedert, Ralf and Hees, J\"{o}rn and Dengel, Andreas and Buscher, Georg},
title = {A Robust Realtime Reading-Skimming Classifier},
year = {2012},
isbn = {9781450312219},
publisher = {Association for Computing Machinery},
address = {New York, NY, USA},
url = {https://doi.org/10.1145/2168556.2168575},
doi = {10.1145/2168556.2168575},
booktitle = {Proceedings of the Symposium on Eye Tracking Research and Applications},
pages = {123–130},
numpages = {8},
location = {Santa Barbara, California},
series = {ETRA '12}
}

@article{sumer2021fakenewsperception,
author = {Ömer Sümer and Efe Bozkir and Thomas Kübler and Sven Grüner and Sonja Utz and Enkelejda Kasneci},
  title = {FakeNewsPerception: An eye movement dataset on the perceived believability of news stories},
journal = {Data in Brief},
volume = {35},
pages = {106909},
year = {2021},
issn = {2352-3409},
doi = {https://doi.org/10.1016/j.dib.2021.106909},
url = {https://www.sciencedirect.com/science/article/pii/S2352340921001931}
}

@article{rayner1998eye,
  title={Eye movements in reading and information processing: 20 years of research.},
  author={Rayner, Keith},
  journal={Psychological bulletin},
  volume={124},
  number={3},
  pages={372},
  year={1998},
  publisher={American Psychological Association},
  url={https://doi.org/10.1037/0033-2909.124.3.372}
}

@article{meziere2023using,
  author = {Mézière, Diane C. and Yu, Lili and Reichle, Erik D. and von der Malsburg, Titus and McArthur, Genevieve},
title = {Using Eye-Tracking Measures to Predict Reading Comprehension},
journal = {Reading Research Quarterly},
volume = {58},
number = {3},
pages = {425-449},
doi = {https://doi.org/10.1002/rrq.498},
url = {https://ila.onlinelibrary.wiley.com/doi/abs/10.1002/rrq.498},
year = {2023}
}

@incollection{just2018using,
  title={Using eye fixations to study reading comprehension},
  author={Just, Marcel Adam and Carpenter, Patricia A},
  booktitle={New methods in reading comprehension research},
  pages={151--182},
  year={2018},
  publisher={Routledge},
  doi={10.4324/9780429505379-8}
}

@inproceedings{papoutsaki2016webgazer,
  author = {Papoutsaki, Alexandra and Sangkloy, Patsorn and Laskey, James and Daskalova, Nediyana and Huang, Jeff and Hays, James},
title = {Webgazer: Scalable Webcam Eye Tracking Using User Interactions},
year = {2016},
isbn = {9781577357704},
publisher = {AAAI Press},
booktitle = {Proceedings of the Twenty-Fifth International Joint Conference on Artificial Intelligence},
pages = {3839–3845},
numpages = {7},
location = {New York, New York, USA},
series = {IJCAI'16},
url={https://dl.acm.org/doi/10.5555/3061053.3061156},

}

@article{reichle2012using,
  title={Using EZ Reader to simulate eye movements in nonreading tasks: A unified framework for understanding the eye--mind link.},
  author={Reichle, Erik D and Pollatsek, Alexander and Rayner, Keith},
  journal={Psychological review},
  volume={119},
  number={1},
  pages={155},
  year={2012},
  publisher={American Psychological Association},
  url={https://psycnet.apa.org/doi/10.1037/a0026473}
}

@inproceedings{Biedert2010,
author = {Biedert, Ralf and Buscher, Georg and Schwarz, Sven and M\"{o}ller, Manuel and Dengel, Andreas and Lottermann, Thomas},
title = {The text 2.0 framework: writing web-based gaze-controlled realtime applications quickly and easily},
year = {2010},
isbn = {9781605589992},
publisher = {Association for Computing Machinery},
address = {New York, NY, USA},
url = {https://doi.org/10.1145/2002333.2002351},
doi = {10.1145/2002333.2002351},
booktitle = {Proceedings of the 2010 Workshop on Eye Gaze in Intelligent Human Machine Interaction},
pages = {114–117},
numpages = {4},
location = {Hong Kong, China},
series = {EGIHMI '10}
}

\end{document}